# Using quantum computing models for graviton communication/information processing in cosmological evolution

A.W. Beckwith
*abeckwith@UH.edu*

## Abstract

In 2001 Seth Lloyd presented a radical critique of how to model the evolution of the universe as a quantum computer. This modeling allows one to reconcile the existence of a huge burst of gravitons with the fact that cosmological CMB is limited by a red shift of about z = 1100. We also discuss difference in values of the upper bound of the cosmological constant between a huge upper bound predicated upon temperature, and the much lower bound predicted by Barvinsky et al in late 2006 with the difference in values in energy input into relic graviton production. Among other things, this difference scaled to Planck's constant accounts for how to introduce quantization via a shift in values of the Hartle-Hawking wave function from a lower value of nearly zero to one which is set via an upper bound of the Planck's constant of the order of 360 times the square of the Planck's mass. It also reconciles the existence of quantization at the onset of the big bang with a requirement that gravitons, so produced interact for up to 1000 years after the big bang itself with ordinary matter, before the red shift limit of about z = 1100, which necessitates a far larger volume of space than what we would use, i.e. the normal Planck length values of 10 to the minus 35 centimeters in length normally associated with the onset of inflation.

## INTRODUCTION

In 2001 Seth Lloyd presented how to model the evolution of the universe as a quantum computer, which allowed him to specify a relationship between presumed entropy levels, and the number of computational bits an evolving universe, as a computational device can 'perform' in its evolution from the big bang state. We will build upon this formalism, with additional developments in the entropy density, i.e. having that vary within the Z = 1100 red shift barrier regime up to the big bang, while assuming up to the Z = 1100 spatial enclosure an approximately constant overall entropy value, which subsequently de creases as we undergo transition from a matter dominated regime for the scalar expansion coefficient, to accelerated values of expansion due to the influence of a dramatically reduced value of the cosmological constant. In doing so, we see evidence of a pitch fork bifurcation of the scalar field value, with a hint of why this could be modeled as being a short term phenomena, as well as how the influx of initial graviton based energy density leading to a high graviton frequency for gravitons. This high graviton frequency could be evidence of a form of 'cosmological' communication of information processing within space time, reconciling the intensity of the big bang event with the red shift barrier of Z = 1100, which previously has not been linked with one another via known physical arguments known to the presenter.

## PRELIMINARY ANALOGY WITH REGARDS TO SETH LLOYDS UNIVERSE AS A QUANTUM COMPUTER PAPER

We make use of the formula given by Seth Lloyd in arXIV quant-ph/0110141 vol 1 24 Oct 2001 which related the number of operations the 'Universe' can 'compute' during its evolution. Seth Lloyd uses the idea he attributed to Landauer to the effect that the universe is a physical system which has information being processed over its evolutionary history. Lloyd also makes reference to a prior paper where he attributes an upper bound to the permitted speed a physical system can have in performing operations in lieu of the Margolis/ Levitin theorem, with a quantum mechanically given upper limit value (assuming E is the average energy of the system above a ground state value), obtaining a **first limit** with regards to a quantum mechanical average energy bound value of

$$[\# operations / \sec] \leq 2E/\pi \hbar \qquad (1)$$

The **second limit** to this number of operations is strictly linked to entropy due to considerations as to limits to memory space. via what Lloyd writes as

$$[\#operations] \leq S(entropy)/(k_B \cdot \ln 2) \qquad (2)$$

The **third limit** as to strict considerations as to a matter dominated universe relates number of allowed computations / operations within a volume for the alleged space of a universe, making the identification of this space time volume as $c^3 \cdot t^3$, with $c$ the speed of light, and $t$ an alleged time/ age for the universe . We further identify $E(energy) \sim \rho \cdot c^2$, with $\rho$ as the density of matter, and $\rho \cdot c^2$ as the energy density/ unit volume. This leads to

$$[\#operations/\sec] \leq \rho \cdot c^2 \times c^3 \cdot t^3 \qquad (3)$$

We then can write this, if we identify $\rho \sim 10^{-27} kil/meter^3$ and time as approximately $t \sim 10^{10}$ years as leading to a present upper bound of

$$[\#operations] \approx \rho \cdot c^5 \cdot t^4 \leq 10^{120} \qquad (4)$$

Seth Lloyd further refines this to read as follows

$$\#operations = \frac{4E}{\hbar} \cdot \left(t_1 - \sqrt{t_1 t_0}\right) \approx \left(t_{Final}/t_P\right) \leq 10^{120} \qquad (5)$$

We assume that $t_1$ = final time of physical evolution, whereas $t_0 = t_P \sim 10^{-43}$ seconds and that we can set an energy input via assuming in early universe conditions that $N^+ \neq \varepsilon^+ \ll 1$, and $0 < N^+ < 1$, so that we are looking at a graviton burst supplied energy value along the lines of

$$E = (V_{4-Dim}) \cdot \left[\rho_{Vac} = \frac{\Lambda}{8\pi G}\right] \sim N^+ \cdot \left[\rho_{graviton} \cdot V_{4-vol} \approx \hbar \cdot \omega_{graviton}\right] \qquad (6)$$

Furthermore, if we use the assumption that the temperature is within the given range of $T \approx 10^{32} - 10^{29}$ Kelvin initially, we have that a Hubble parameter defined along the route specified by Seth Lloyd. This is in lieu of time $t = 1/H$, a horizon distance defined as $\approx c/H$, a total energy value within the horizon as

$$\text{Energy (within the horizon)} \approx \rho_C \cdot c^3 / (H^4 \cdot \hbar) \approx 1/(t_P^2 \cdot H) \qquad (7)$$

And this for a Horizon parameter Seth Lloyd defines as

$$H = \sqrt{8\pi G \cdot [\rho_{crit}]/3 \cdot c^2} \qquad (8)$$

And a early universe

$$\rho_{crit} \sim \rho_{graviton} \sim \hbar \cdot \omega_{graviton}/V_{4-Vol} \qquad (9)$$

Then

$$\#operations \approx 1/[t_P^2 \cdot H] \approx \sqrt{V_{4-Vol}} \cdot t_P^{-2}/\sqrt{[8\pi G \hbar \omega_{graviton}/3c^2]} \approx [3\ln 2/4]^{4/3} \cdot [S_{Entrophy}/k_B \ln 2]^{4/3} \qquad (10)$$

# II. INVESTIGATION OF ENTROPY WITHIN THE Z=1100 RED SHIFT CMB BARRIER\

So far we have argued as to the existence of a high level of entropy within what could be a large volume of space relative to the initial volume present in vacuum nucleation of an initially very high energy density state of matter-energy. We shall endeavor to give more specifics as to the relationship between entropy density, overall entropy, changes in volume, as well as changes in time which we believe govern the spatial regime in which gravitons are initially coupled to matter. Key to this is asserting that the entropy density is scaled in early Universe conditions as an initial value of entropy density at a proper time proportional to Planck's time interval of about ten to the minus 43 seconds in magnitude times the ration of initial proper time, over proper time at a later stage in cosmological evolution. . Whereas the overall entropy is the entropy density, times a space time volume specified at a given proper time. For those who wish to understand what we are referring to, we define first a de facto distance to any comoving observer, where $D_G(t_{present})$ is the distance $D_{now}$ to galaxy G *now*, while a(t) is a universal scale factor that applies to all comoving objects. From its definition we see that $a(t_{present}) = 1$ so we get a generalized distance relationship

$$D_G(t) = a(t) * D_G(t_{present}) \tag{11}$$

This is assuming a large red shift value, where we define the red shift value Z via

$$1+z = \sqrt{(1+v/c)/(1-v/c)} \tag{12}$$

The proper time so referred to, on small scales, is based on the concept of volume, while for large scales the usual definition of length is applied. This is given in arXiv:gr-qc/0102088, which we refer to in our bibliography. Needless to say for area/volume about the big bang, proper time is proportional to the Planck time interval $\propto 10^{-43}$ seconds. This leads to us following scaling of entropy density, as given by Seibert (and Bjorken in 1983) in 1991 as, where we define $\tau_0$ as an initial (at the point of nucleation of a vacuum state) proper time value, so we write entropy density for proper times $\tau$ greater than $\tau_0$

$$s(\tau) = s(\tau_0) \cdot \left[\frac{\tau_0}{\tau}\right] \tag{13}$$

If we make a relationship between entropy density, and entropy itself via $S(\tau) = s(\tau_0) \cdot [V_{4-\dim}] \cdot \left[\frac{\tau_0}{\tau}\right]$, we can define the onset of early stages of entropy as growing up to a time Planck time interval $\propto 10^{-43}$ seconds via Smibert's formula of

$$S(t) = k \cdot \sigma \cdot t^2, \quad 0 < t < [\lambda/2c] \approx t_P \sim 10^{-43} \text{ seconds} \tag{14}$$

This would correspond to the Loop quantum gravity insertion of thermal energy into a present universes space time continuum with an initial relic graviton producing burst initiated by a cosmological 'constant' $\Lambda_{Park} \approx \infty \xrightarrow[Axion\ mass \to 0]{} \Lambda_{Barvinsky} \approx 360 \cdot m_P^2$, with the lower value signifying a release of relic gravitons, and this when we are setting $m_P$ as the Planck mass, i.e. the mass of a black hole of 'radius' on the order of magnitude of Planck length $l_P \sim 10^{-35}$ centimeters in width. Hereafter, we are then setting the entropy as scaling as

$$S(t) = \langle s \rangle_{net} \cdot [V_{4-\dim}(t)], \quad t_P \sim 10^{-43} < t < t\left(at \quad z \cong 1100\right) \tag{15}$$

This is assuming that the $\langle s \rangle_{net}$ is an average value of entropy density which would be relatively constant in the aftermath of the big bang up to the red shift barrier of Z of the order of 1100. Then afterwards, we could expect a rough scaling of entropy density according to

$$S(\tau) = s(\tau_0) \cdot [V_{4-\dim}(time)] \cdot \left[\frac{\tau_0}{\tau}\right] \qquad (16)$$

We wish now that we have described how entropy behaves via scaling arguments look at the physical inputs into this problem Note, we are making one specific identification here. That the initial growth of graviton based entropy density is in tandem with the growth of free energy with increasing temperatures in the onset of a vacuum state according to

$$\text{Free energy} \sim -\pi^2 T^4/90 \approx -s(T)^2 \qquad (17)$$

We obtained this value of free energy by associating Eqn. (17) with the free energy of a massless spin zero boson, or minus the pressure of a 'spin zero boson' state, and $\left[\pi^2 T^4/30\right]$ as the energy density of a spin zero boson gas, which can be read off as part of a finite temperature one loop full potential $V_T(\phi_C)$ for high temperature values of a scalar field given by (assuming that $V(\phi_C)$ is a one loop effective potential) when we are looking at an a early universe scalar field model of inflation for which we are looking at high temperature symmetry restoration , where we look at a critical value of the scalar field at about the time of vacuum nucleation, we call $\phi_C$ at time $t \approx 10^{-43}$ seconds.

$$V_T(\phi_C) = V(\phi_C) + \frac{\lambda}{8} T^2 \phi_C^2 - s(T^2) + \ldots \qquad (18)$$

This leads to us now considering what physical inputs should be made into the parameters of our entropy/ number of computations upper limits as specified by the given equations written up in this document

## III. INTRODUCTION TO PHYSICAL INPUTS INTO THIS PROBLEM

First of all we need to consider if there is an inherent fluctuation in early universe cosmology which is linked to a vacuum state nucleating out of 'nothing'. The answer we have is yes and no.

The vacuum fluctuation leads to production of a dark energy density which we can state is initially due to contributions from an axion wall, which is dissolved during the inflationary era. What we will be doing is to reconcile how that wall was dissolved in early universe cosmology with quantum gravity models, brane world models, and Weinberg's prediction (published as of 1972) of a threshold of 10 to the 32 power Kelvin for which quantum effects become dominant in quantum gravity models. All of this leads up to conditions in which we can expect relic graviton production which could account for the presence of strong gravitational fields in the onset of Guth style inflation, would be in line with Penrose's predictions via the Jeans inequality as to low temperature, low entropy conditions for pre inflationary cosmology.

It is noteworthy that Barvinsky et al in late 2006 recently predicted a range of values of four dimensional Planck's constant between upper and lower bounds. I.e. this is a way to incorporate the existence of a cosmological constant at about a Planck's time $t_P$ with the formation of scale factors which permit the existence of definable space time metrics. A good argument can be made that prior to Planck's time $t_P$ that conventional space time metrics, even those adapting to strongly curved space do not apply. Park et al (REF) predict an upper range of cosmological constant values far in excess of Barvinsky's prediction, and we explain the difference in terms of a thermal/ vacuum energy input into graviton production. We will henceforth investigate how this would affect the emergence of an initial state for the scale factor, in the cases where the cosmological constant is first a lower bound, and then where the cosmological constant parameter is grows far larger.

.In order to do this, we first examine how the Friedman equation gives us an evolution of the scale factor a(t) , in two cases, .Case one will be with a constant cosmological constant. And case two will be when the cosmological constant was far larger than it is today.

Very large cosmological constant will lead to a road map to solving the land scape problem. I.e. $\Lambda_{\max}(Barvinsky) = 3m_P^2/2B = 360 m_P^2$ as a peak value, after graviton production would lead to a Hartle-Hawking's universe wave function of the form

$$\psi_{HH}\big|_{Barvinsky} \approx \exp(-S_E) = \exp(3 \cdot \pi / 2 \cdot G\Lambda) \neq 0 \tag{19}$$

Conversely Parks values for a nearly infinite cosmological constant parameter, due to high temperatures would lead to, prior to graviton production

$$\psi_{HH}\big|_{Park} \approx \exp(-S_E) = \exp(3 \cdot \pi / 2 \cdot G\Lambda) \xrightarrow[T \to \infty]{} 0 \tag{20}$$

This allows us to make in roads into a solution to the cosmological land-scape problem discussed by Guth in 2003 at the Kalvi institute in UC Santa Barbara. I.e. why have $10^{1000}$ or so independent vacuum states as predicted by String theory?

## II STATEMENT OF THE GENERAL PROBLEM WE ARE INVESTIGATING

If one looks at the range of allowed upper bounds of the cosmological constant, we have that the difference between what Barvinsky et al in late 2006 predicted, and Park's upper limit as of 2003, based upon thermal input strongly hints that a phase transition is occurring at or before Planck's time . This allows for a brief interlude of quintessence

Begin with assuming that the absolute value of the five dimensional cosmological 'constant' parameter is inversely related to temperature, i.e.

$$|\Lambda_{5-\dim}| \propto c_1 \cdot (1/T) \tag{21}$$

As opposed to working with the more traditional four dimensional version of the same. Those wishing to get specific values of the constants $c_1$ and $c_2$ are referred to look at the Beckwith (2006) contribution to STAIF new frontiers section, and also Beckwith's gr-qc article mentioned in the references, which phrased the release of gravitons in terms of

$$\Lambda_{4-\dim} \propto c_2 \cdot T \tag{22}$$

We should note that this is assuming that a release in gravitons occurs which leads to the removal of graviton energy stored contributions to this cosmological parameter

$$\Lambda_{4-\dim} \propto c_2 \cdot T \xrightarrow[graviton-production]{} 360 \cdot m_P^2 << c_2 \cdot \left[T \approx 10^{32} K\right] \tag{23}$$

Needless to say, right after the gravitons are released one still observes a drop off of temperature contributions to the cosmological constant .Then we can write, for small time values $t \approx \delta^1 \cdot t_P$, $0 < \delta^1 \leq 1$ and for temperatures sharply lower than $T \approx 10^{12} Kelvin$

$$\left(\frac{\Lambda}{|\Lambda_5|} - 1\right) \approx O\left(\frac{1}{n}\right) \sim \text{ To the order of } (1/n) \tag{24}$$

for quantum effects to be dominant in cosmology, with a value of critical energy we will use in setting a template for relic graviton production later on.

$$E_{critical} \equiv 1.22 \times 10^{28} eV \qquad (25)$$

This is presupposing that we have a working cosmology which actually gets to such temperatures at the instance of quantum nucleation of a new universe.

## III. GRAVITON POWER BURST/ WHERE DID THE MISSING CONTRIBUTIONS TO THE COSMOLOGICAL 'CONSTANT 'PARAMETER GO?

To do this, we need to refer to a power spectrum value which can be associated with the emission of a graviton. Fortunately, the literature contains a working expression as to power generation for a graviton being produced for a rod spinning at a frequency per second $\omega$, due to Fontana (2005) at a STAIF new frontiers meeting, which reportedly gives for a rod of length $\hat{L}$ and of mass m a formula for graviton production power,

$$P(power) = 2 \cdot \frac{m_{graviton}^2 \cdot \hat{L}^4 \cdot \omega_{net}^6}{45 \cdot (c^5 \cdot G)} \qquad (26)$$

Note here that we need to say something about the contribution of frequency needs to be understood as a mechanical analogue to the brute mechanics of graviton production. We can view the frequency $\omega_{net}$ as an input from an energy value, with graviton production number (in terms of energy) as given approximately via an integration of Eqn. (27) below, $\hat{L} \propto l_P$ mass $m_{graviton} \propto 10^{-60} kg$. It also depends upon a huge number of relic gravitons being produced, due to the temperature variation so proposed.

$$\langle n(\omega) \rangle = \frac{1}{\omega(net\ value)} \int_{\omega 1}^{\omega 2} \frac{\omega^2 d\omega}{\pi^2} \cdot \left[ \exp\left(\frac{2 \cdot \pi \cdot \hbar \cdot \omega}{\bar{k}T}\right) - 1 \right]^{-1} \qquad (27)$$

Thus, one can set a normalized 'energy input 'as $E_{eff} \equiv \langle n(\omega) \rangle \cdot \omega \equiv \omega_{eff}$ ; with $\hbar\omega \xrightarrow[\hbar \equiv 1]{} \omega \equiv |E_{critical}|$ which leads to the following table of results, with $T^*$ being a heating up value of temperature from a brane world thermal input from a prior universe quantum bounce after a nearly zero degrees Kelvin starting point of the pre inflationary universe condition specified by Carroll.. For the sake of scaling, we will refer to $T^* \sim \frac{1}{3} \times 10^{32}$ Kelvin, with a peak graviton burst happening at the time where quantum gravity becomes a dominant contribution to early universe vacuum energy nucleation. I.e. this phase transition occurs in a very brief instant of cosmological time with the onset of the graviton burst being modeled at times $t << t_P \sim 10^{-43}$ .seconds. The values of N1, to N5 are partly scaled graviton burst values. The tie in of a relic graviton burst so presented with brane world models has been partly explained in the author's STAIF (2006) publication and our description of a link of the sort between a brane world effective potential and eventual Guth style inflation has been partly replicated by Sago, Himenoto, and Sasaki in November 2001 where they assumed a given scalar potential, assuming that *m* is the mass of the bulk scalar field. This permits mixing the false vacuum hypothesis of Coleman in 4 dimensions with brane world theory in five dimensions.

$$V(\phi) = V_0 + \frac{1}{2}m\phi^2 \qquad (28)$$

Their model is in part governed by a restriction of their 5-dimensional metric to be of the form, with $l$ = brane world curvature radius, and $\hat{H}$ H their version of the Hubble parameter

$$dS^2 = dr^2 + (\hat{H} \cdot l)^2 \cdot dS^2_{4-dim} \qquad (29)$$

I.e. if we take $k_5^2$ as being a 5 dimensional gravitational constant

$$\hat{H} = \frac{k_5^2 \cdot V_0}{6} \qquad (30)$$

Our difference with Eqn. (28) is that we are proposing that it is an intermediate step, and not a global picture of the inflation field potential system. However, the paper they present with its focus upon the zero mode contributions to vacuum expectations $\langle \delta\phi^2 \rangle$ on a brane has similarities as to what we did which should be investigated further. The difference between what they did, and our approach is in their value of

$$dS^2_{4-dim} \equiv -dt^2 + \frac{1}{H^2} \cdot \left[\exp(2 \cdot \hat{H} \cdot t)\right] \cdot dx^2 \qquad (31)$$

This assumes one is still working with a modified Gaussian potential all the way through, as seen in Eqn. (11). This is assuming that there exists an effective five dimensional cosmological parameter which is still less than zero, with $\Lambda_5 < 0$, and $|\Lambda_5| > k_5^2 \cdot V_0$ so that

$$\Lambda_{5,eff} = \Lambda_5 + k_5^2 \cdot V_0 < 0 \qquad (32)$$

It is simply a matter of having

$$|m^2| \cdot \phi^2 << V_0 \qquad (33)$$

And of making the following identification

$$\phi_{5-dim} \propto \tilde{\phi}_{4-dim} \equiv \tilde{\phi} \approx \left[\phi - \varphi_{fluctuations}\right]_{4-dim} \qquad (34)$$

With $\varphi_{fluctuations}$ in Eqn. (34) is an equilibrium value of a true vacuum minimum for a chaotic four dimensional quadratic scalar potential for inflationary cosmology. This in the context of the fluctuations having an upper bound of $\tilde{\tilde{\phi}}$ (Here, $\tilde{\tilde{\phi}} \geq \varphi_{fluctuations}$ ). And $\tilde{\phi}_{4-dim} \equiv \tilde{\tilde{\phi}} - \frac{m}{\sqrt{12 \cdot \pi \cdot G}} \cdot t$ , where we use $\tilde{\tilde{\phi}} > \sqrt{\frac{60}{2 \cdot \pi}} M_P \approx 3.1 M_P \equiv 3.1$, with $M_P$ being a Planck mass. This identifies an imbedding structure we will elaborate upon later on. This will in its own way lead us to make sense of a phase transition we will write as a four dimensional embedded structure within the 5 dimensional Sundrum brane world structure and the four dimensional

$$\begin{array}{ll} \tilde{V}_1 & \to \tilde{V}_2 \\ \tilde{\phi}(increase) \leq 2 \cdot \pi & \to \tilde{\phi}(decrease) \leq 2 \cdot \pi \\ t \leq t_P & \to t \geq t_P + \delta \cdot t \end{array} \qquad (35)$$

The potentials $\tilde{V}_1$, and $\tilde{V}_2$ will be described in terms of chaotic inflationary scalar potential system. Here,

$m^2 \approx (1/100) \cdot M_P^2$

$$\tilde{V}_1(\phi) \propto \frac{m_a^2(T)}{2} \cdot \left(1 - \cos(\tilde{\phi})\right) + \frac{m^2}{2} \cdot \left(\tilde{\phi} - \phi^*\right)^2 \qquad - \qquad (36)$$

$$\tilde{V}_2(\phi) \propto \frac{1}{2} \cdot (\tilde{\phi} - \phi_C)^2 \tag{37}$$

The transition from Eqn. (36) above to Eqn. (37) is when we have a relic graviton burst as given below which also is when we have a removal of an axion wall contribution with the axion mass term $m_a(T) \xrightarrow[Temp \to Planck\ temperature]{} \varepsilon^+ \approx$ very small value almost non existent in contribution due to temperature scaling as given below.

$$m_a(T) \cong 0.1 \cdot m_a(T=0) \cdot (\Lambda_{QCD}/T)^{3.7} \tag{38}$$

We assert that we need a five dimensional brane world picture to formulate what configuration the non zero axion mass makes initially to a Sundrum initial compactified 5$^{th}$ dimensional presentation of the action integral as given in Beckwith's (2006) STAIF document. As the contribution to Eqn. (38) vanishes, we see the following graviton burst.

### HOW TO OUTLINE THE EXISTENCE OF A RELIC GRAVITON BURST

| | |
|---|---|
| N1=1.794 E-6 for $Temp = T^*$ | Power = 0 |
| N2=1.133 E-4 for $Temp = 2T^*$ | Power = 0 |
| N3= 7.872 E+21 for $Temp = 3T^*$ | Power = 1.058 E+16 |
| N4= 3.612E+16 for $Temp = 4T^*$ | Power $\cong$ very small value |
| N5= 4.205E-3 for $Temp = 5T^*$ | Power= 0 |

The outcome is that there is a distinct power spike associated with Eqn. 26 and Eqn. 27, which is congruent with a relic graviton burst. Future research objectives will be to configure the conditions via brane world dynamics leading to graviton production. A good working model as to how the cosmological constant changes in this comparison of four and five dimensional cosmological constants is provided in tables given in Beckwith arXIV physics article mentioned in the reference section..

### IV. CONCLUSION

So far, we have tried to reconcile the following.

First is that Brane world models will not permit Akshenkar's quantum bounce. The quantum bounce idea is used to indicate how one can reconcile axion physics with the production of dark matter/dark energy later on in the evolution of the inflationary era where one sees Guth style chaotic inflation for times $t \geq t_P$ and the emergence of dark energy during the inflation era.

In addition is the matter of Sean Carroll, J. Chens paper which pre supposes a low entropy – low temperature pre inflationary state of matter prior to the big bang. How does one ramp up to the high energy values greater than temperatures $10^{12}$ Kelvin during nucleosynthesis? The solution offered is novel and deserves further inquiry and investigation.

Now for future research goals. Looking at Eqn (4) of the second page, we would have a dramatically lowered value for a net range of graviton frequencies if the initial volume of space for graviton production is localized in the regime near the Planck time interval. I.e. we may need to, for information theory reasons go out to the $z \approx 1100$ red shift limit years after the big bang to commence a region of space consistent with Eqn. (4) of the first page, with high net graviton frequencies. This is assuming a large initial cosmological constant.

We have the paradoxical result that we may need a huge influx of gravitons to give the initially low temperature, low entropy initial conditions given by Sean Carroll; the initially low temperature conditions changed by a gigantic cosmological constant, which after graviton production would lead to Barvinsky's lower value of the maxim value of 360 times the square of Planck's mass to give us an answer to the cosmological constant problem. I.e. The Hartle-Hawking wave function comes into being right after graviton production, whereas it is zero initially beforehand. This would lead to a favored state as to the nucleation of the cosmological landscape due to changes in the cosmological constant at or before a Planck's time interval, where we would have $z \sim 10^{26}$, whereas we would have to go out to the region of space where $z \approx 1100$ to be consistent with regards to Seth Lloyds measurement of the computational space time limits of how the universe evolves in time. We assert that quantum computing models as given by Seth Lloyd are providing a good probing of the phase transition we assert between the $z \sim 10^{26}$ regime in the spatial environs of the big bang with the $z \approx 1100$ CMBR barrier. This also will allow us to understand the physics inherent in how we can get phenomenological verification, if possible of experimental observation of gravitons in a way answering T. Rothmans treatment of gravitons, and also allow us to refine Pisen. Chens Sach – Wolfe calculations This would allow us to understand, via an improved quantum computation model necessary and sufficient conditions for the onset of graviton interaction with the generation of a huge gravitation surge with a relic magnetic field. Note, P. Chen value of that magnetic field was $B_{relic} \leq 10^{11}$ Gauss, for graviton interactions with early universe 'matter' at about one minute of time after the big bang. This is in contrast to the value of that initial magnetic field as allegedly being of the form $B_{today} \leq 10^{-7}$ Gauss as measurable via space craft instrumentation in today's astrophysical data collection milieu What Pisen Chen gives us in his 1994 paper is a way to tie in the purported value of the relic magnetic field, as proportional to temperature fluctuations, as given by the following upper bound inequality. This is for the Planck 'constant value $H_1$ left as an 'open parameter' but understood as being set at the ONSET of inflation, $\omega_0$ being the initial frequency of a graviton production process, and

$$\delta\Omega_{EM} \sim \frac{B_*^2}{8\pi} \cdot \frac{1}{\rho_C^*} \text{ as well as } T_{present} \sim 2.7^0 K$$

$$\left[\frac{\sinh^2(\omega_0/T_{present})}{(\omega_0/T_{present})}\right] \cdot \left[\frac{4\pi^4}{15}\right] \cdot \left[\frac{10^{-29} \times T_{present}}{H_1}\right]^4 \cdot \delta\Omega_{EM}^* \leq \left\langle\frac{\delta T}{T}\right\rangle \tag{39}$$

If the ratio of $\omega_0$ to $T_{present} \sim 2.7^0 K$ is large, this means that we observe a de facto Sach-Wolfe phenomena via

$$\delta\Omega_{EM} \sim \frac{B_*^2}{8\pi} \cdot \frac{1}{\rho_C^*} \sim \left\langle\frac{\delta T}{T}\right\rangle \tag{40}$$

We are hoping that improvements upon the estimation of the phase transition as can be glimpsed by understanding further how entropy plays a role in the phase transition indicated above will allow us the means to refine both Eqn. 39 and 40, in ways permitting a good way to make the Sach-Wolfe effect a tool to probe relic graviton production .This also would be a way to refine current models of the relationship between scalar and tensor models of early universe fluctuations as mentioned by Lukash. We assert all of this is possible if we use Seth Lloyd's quantum computational model of the universe judicially and with common sense.

# NOMENCLATURE

$M_P \approx 2.176 \times 10^{-8} [kg]$  $\quad t_P = \sqrt{\hbar \cdot G / c^5}$

$G = 6.67300 \times 10^{-11} \ m^3 / kg \ s^2$  $\quad g = \det(g_{ab})$

$r_s = 2Gm/c^2$  $\quad \bar{\bar{R}}$ – Ricci scalar

$\lambda_C = 2 \cdot \pi \cdot \hbar / m \cdot c$